# Calculation of AC losses in stacks and coils made of second generation high temperature superconducting tapes for large scale applications

**Victor M. R. Zermeno**[1,4,a], **Asger B. Abrahamsen**[2], **Nenad Mijatovic**[3], **Bogi B. Jensen**[3] and **Mads P. Sørensen**[4]

[1]Karlsruhe Institute of Technology, P.O. Box 3640, 76021, Karlsruhe, Germany
[2]Department of Wind Energy, Technical University of Denmark, Frederiksborgvej 399, 4000 Roskilde, Denmark
[3]Department of Electrical Engineering, Technical University of Denmark, Anker Engelunds Vej 1, 2800 Kgs. Lyngby
[4]Department of Mathematics, Technical University of Denmark, Anker Engelunds Vej 1, 2800 Kgs. Lyngby

**Abstract.** A homogenization method to model a stack of second generation (2G) High Temperature Superconducting (HTS) tapes under AC applied transport current or magnetic field has been obtained. The idea is to find an anisotropic bulk equivalent for the stack, such that the geometrical layout of the internal alternating structures of insulating, metallic, superconducting and substrate layers is "washed" out while keeping the overall electromagnetic behavior of the original stack. We disregard assumptions upon the shape of the critical region and use a power law E-J relationship allowing for overcritical current densities to be considered. The method presented here allows for a computational speedup factor of up to 2 orders of magnitude when compared to full 2-D simulations taking into account the actual dimensions of the stacks without compromising accuracy.

## I. INTRODUCTION

Second generation (2G) HTS coated conductors have already come to play an important role in a large number of power applications. Nowadays, superconducting cables such as the increasingly popular Roebel are being used for their high current capacity [1,2]. In the same manner, motors, generators, transformers and large magnets are designed and/or built taking advantage of the high magnetic field achieved by superconducting coils or windings in compact designs. Although some of these devices are designed so that their superconducting elements do not experience AC electromagnetic fields, hysteretic losses are expected during start up, turn off and other transient operations. Furthermore, transformers, asynchronous rotating machinery, and cables carrying AC currents are inherently burdened by hysteretic losses. Understanding and calculating them is fundamental for performance evaluation and design. From the modeling point of view, all the aforementioned applications rely on a similar basic unit: a stack of superconducting tapes. The cross section of racetrack coils, as the ones conforming radial flux electric motors or generators, can be modeled as a couple of stacks transporting current in opposite directions. Circular coils can be treated in a similar way by using cylindrical coordinates. Finally, Roebel cables can be modeled as two parallel stacks of tapes where all the strands carry the same net current [1]. Hence, calculation of AC losses in large scale HTS devices

---
[a] Electronic mail: victor.zermeno@kit.edu

can be reduced to computing the equivalent eddy currents problem for stacks of tapes.

The large aspect ratio of the thin films in 2G HTS coated conductors shows the multiscale nature of the layout: thickness and width are in different spatial scales. This later problem was already addressed in an earlier work of ours [3] where structured meshes were used to achieve a computational speedup of 2 to 3 orders of magnitude. Modeling and simulation of large stacks of 2G HTS coated conductors under AC conditions has already been the subject of study by means of: integral equations using a thin conductor approximation for the cases of infinite stacks, periodic arrays or a couple of conductors [4]; direct integration also using a thin conductor approximation for the cases of infinite bifilar stacks [5]; quasi-variational inequalities [6] detailing the actual layout of just a few conductors; partial differential equations describing stacks of up to 100 conductors [3,7]; and anisotropic homogenous-medium approximations [8–10] used for arbitrarily large stacks.

Regarding the anisotropic homogenous-medium approximations, three important works are worth mentioning. In a pioneering work, Clem et al. [8] adhered to the following assumptions: 1) The critical current density $J_c$ of the superconducting layers is constant. 2) The magnetic field is parallel to the tape surface inside the subcritical region of the equivalent bulk, hence for that region it can be assumed that $\partial J/\partial x = 0$. 3) The boundary between the critical and subcritical zones can be approximated with a straight line perpendicular to the tape surface. In a further work by Yuan et al. [9] the first assumption is discarded by allowing a Kim like model for the critical current density $J_c(B)$ dependence. Although the second assumption is kept, the third is improved by using parabolas to fit the boundary between the critical and subcritical zones. More recently, a further improved model was presented by Prigozhin and Sokolovsky [10]. Their model for the anisotropic bulk limit, based on a quasi-variational inequality formulation, does not rely on any assumptions for the shape of boundaries separating the critical and subcritical zones in the stack. However, this formulation is based upon the critical state model using a Kim-like $J_c(B)$ dependence. Hence, it assumes a zero electric field E for all subcritical regions and does not allow for considering overcritical local currents. In the present work, a further generalization for the anisotropic bulk model is described where none of the aforementioned assumptions is considered. The treatment is based upon the widely used H-formulation using edge elements [11–14] and a power law to describe the $\boldsymbol{E} - \boldsymbol{J}$ relationship ($|\boldsymbol{E}| \propto |\boldsymbol{J}|^n$). A scaled Kim-like model is used to characterize the $J_c(B)$ dependence. The model for the homogenized stack is compared to a fully featured stack of tapes to evaluate its accuracy for both cases of transport current and perpendicularly applied magnetic field. All calculations were performed using the commercially available Finite Element Method (FEM) software package COMSOL Multiphysics [15].

## II. MODELING STRATEGY
### A. H-formulation

Early uses of a formulation in the magnetic field H for modeling superconductors can be traced back to Kajikawa et al. [12] and to Pecher et al. [11]. The latter, already used edge elements for performing numerical simulations. However, their treatment does separate the self and applied contributions of the magnetic field. Strategies involving only the physical magnetic field H appear in Hong et al. [13] and Brambilla et al. [14]. A further study made by Nguyen et al. [16], allowed extending the model to consider materials with nonlinear B − H relations. The formulation described in the present work will correspond to the one used by Brambilla et al. [14] as integral constraints are used to impose individual currents to different conductors, and no separation is made between the self and external fields. In this work the H-formulation has been chosen due to its ease and simplicity for implementation. As described in the appendix, use of zeroth-order edge elements allows for a direct computation of the current density from the calculated magnetic field components without the need for additional numerical differentiation, hence providing with a high degree of accuracy. Furthermore, use of structured meshes[3] provides with an easy method for meshing thin or rectangular shaped domains as the ones used in this work.

To model a stack or a coil, its cross section is considered and assumed to be a bundle of parallel conductors composed of both normal and superconducting materials. The computational domain of interest $\Omega$ is shown in FIG. 1. If the conductors are coupled at the ends, transport current can be imposed by means of a Dirichlet condition at the domain boundary $\partial\Omega$, said boundary is typically set at a distance of 5 to 10 times the maximum cross-sectional diameter of the conductors bundle. However, for the more general case of a given current being enforced in each conductor, said Dirichlet condition alone does not suffice. In general, for a group of $n_c$ parallel conductors – each carrying a prescribed current $I_k(t)$, $k \in \{1,2,...,n_c\}$ – one integral constraint per conductor ensures the transport current requirement is met.

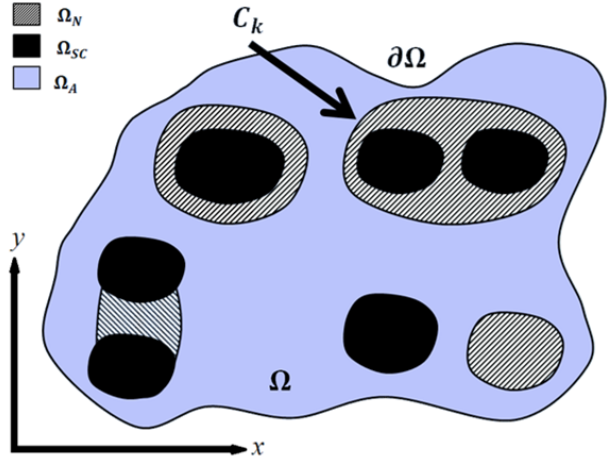

FIG. 1. The computational domain $\Omega = \Omega_A \cup \Omega_N \cup \Omega_{SC}$, represents the cross-section of an infinitely long bundle of conductors. The superconducting regions are denoted by $\Omega_{SC}$, the normal conductors by $\Omega_N$ and the air or insulation by $\Omega_A$. Neither $\Omega_{SC}$ nor $\Omega_N$ need to be connected. The boundary of the computational domain is denoted by $\partial\Omega$. The arrow points to the $k^{\text{th}}$ conductor which carries a net current $I_k(t)$.

Here, $J = \hat{e}_3 \cdot \nabla \times H$ is the component of the current density in the direction out of the $x$-$y$ plane and $I_k(t)$, the net transport current in the conductor $C_k$. If each tape carries the same current $I(t)$ as for the case of coils or Roebel cables, then $\forall k \in \{1,2,...,n_c\}$, $I_k(t) = I(t)$. Finally, the eddy currents problem is reduced to find the magnetic field $H$ such that [17]:

$$\nabla \times \rho \nabla \times H = -\mu \frac{\partial H}{\partial t} \quad in\ \Omega, \tag{1}$$

$$\boldsymbol{H} = \boldsymbol{H}_{self} + \boldsymbol{H}_{ext} \quad on\ \partial\Omega, \tag{2}$$

$$\boldsymbol{H}_{t=0} = \boldsymbol{H}_0 | \nabla \cdot (\mu\ \boldsymbol{H}_0) = 0 \tag{3}$$

and

$$I_k(t) = \int_{C_k} \boldsymbol{J}\ dA \quad \forall k \in \{1,2,\dots,n_c\}. \tag{4}$$

Values for $\boldsymbol{J}$, $\boldsymbol{E}$ and $\boldsymbol{B}$ can be obtained from $\boldsymbol{H}$ using Ampere's law $\boldsymbol{J} = \nabla \times \boldsymbol{H}$ and the constitutive equations $\boldsymbol{E} = \rho \boldsymbol{J}$ and $\boldsymbol{B} = \mu \boldsymbol{H}$, respectively. Here, the resistivity of the superconducting material is modeled by a power law dependence on the current density $\boldsymbol{J}$ as follows:

$$\rho_{HTS} = \frac{E_c}{J_c} \left|\frac{\boldsymbol{J}}{J_c}\right|^{n-1}. \tag{5}$$

Here $E_c = 1\mu V/cm$ is the electric field when the critical current density $J_c$ is reached and $n$ is the exponent in the $\boldsymbol{E} - \boldsymbol{J}$ relationship. Although, in principle any initial condition $\boldsymbol{H}_0$ fulfilling $\nabla \cdot (\mu\ \boldsymbol{H}_0) = 0$ can be used, for simplicity $\boldsymbol{H}_0 = 0$ was chosen. Finally, calculation of instantaneous AC losses (in W/m) can be achieved by evaluating of the following integral:

$$\xi = \int_\Omega \boldsymbol{E} \cdot \boldsymbol{J} d\Omega. \tag{6}$$

For periodic driving signals (either applied magnetic field or transport current), average hysteretic losses (in W/m) can also be calculated by means of the following integral:

$$Q = \frac{1}{T}\int_T^{2T} dt \int_\Omega \boldsymbol{E} \cdot \boldsymbol{J} d\Omega. \tag{7}$$

Transient phenomena are expected due to the zero initial condition chosen. However, given the hysteretic nature of the losses, this transient will fade out once the external field or driving current has reached a maximum value. Hence, time integration takes place only in the second cycle to ensure this maximum has been reached. In the particular case of sinusoidal excitations, and for the sake of computational speed, this integral can be changed to:

$$Q = \frac{2}{T}\int_{T/2}^{T} dt \int_\Omega \boldsymbol{E} \cdot \boldsymbol{J} d\Omega. \tag{8}$$

### B.    Homogenization

Consider the vertical stack of 2G HTS tapes as described in **FIG. 2**. Although in a real application, the separation between the tapes $D$, could be a packing quality parameter changing throughout the stack, it will be considered constant in this study. The stack is presented as a periodic linear array of unit cells. Up to the μm scale, this layout corresponds to a 2G HTS coated conductor manufactured with the ion beam assisted deposition technique (IBAD) [18]. Then, a unit cell will be composed of layers of copper, silver, YBa$_2$Cu$_3$O$_{6+x}$ (YBCO) superconductor, substrate and the air/insulation separating it from the next tape. In order to consider an anisotropic bulk model, the actual topological features of the tapes are "washed out". Therefore, material parameters have to be modified accordingly.

Resistivity values of air and normal conductors are several orders of magnitude bigger than those of superconductors in the mixed state. For this reason, in the homogenization process only the superconducting material's volume fraction will be considered.

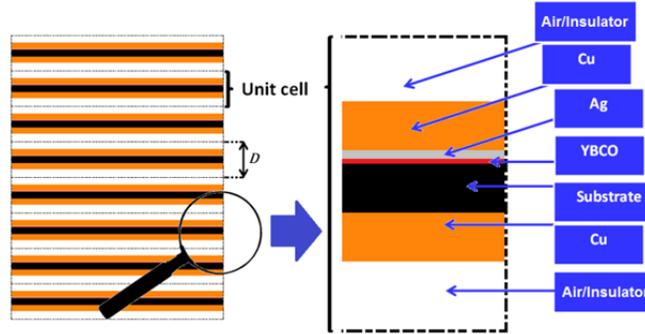

FIG. 2. Stack of coated conductor superconducting tapes as a periodic array of unit cells with height $D$. The zoomed image (*right*) shows the internal layered structure of each unit cell.

To take into account the $J_c(\boldsymbol{B})$ dependence, a Kim like model is considered [19]. For the homogenous bulk, the equivalent engineering critical current density $J_{c,Eq}(\boldsymbol{B}) = J_c(\boldsymbol{B}) f_{HTS}$ is used; here $f_{HTS}$ is the volume fraction of the superconducting material per unit cell. In what follows and for ease, $J_c$ alone will refer to the equivalent engineering critical current density $J_{c,Eq}(\boldsymbol{B})$ in the case of a homogenized stack and to $J_c(\boldsymbol{B})$ in the case of the superconducting layers of a tape. Tapes manufactured using the IBAD technique, do not employ magnetic substrates. Therefore, the relative permeability of the various layers is considered equal to one.

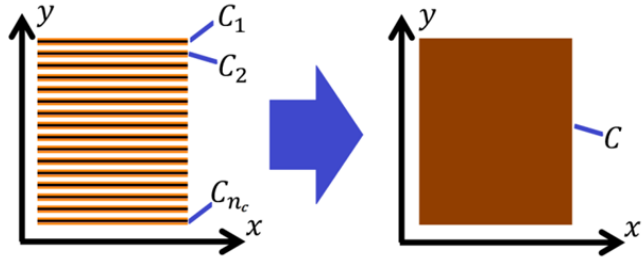

FIG. 3. Homogenization a stack of tapes. The labels $C_1, C_2, \ldots, C_{n_c}$ and $C$ denote each of the tapes in the actual stack (*left*), and the homogeneous-medium equivalent (*right*), respectively.

To model the stack of $n_c$ conductors as shown in the left side of FIG. 4, where each conductor carries a prescribed current, one constraint (4) per conductor ensures the requirement is met. Numerically, this is implemented by the introduction of a set of $n_c$ Lagrange multipliers, one for each constraint.

However, the use of a homogenized domain as shown in the right half of FIG. 3, does not allow for implementing the described constraints. Hence, a new constraint will be required to assure the intended transport current. In the limiting case of a tightly packed stack composed by infinitely thin conductors, this condition can be expressed as [8]:

$$K(y,t) = \int_C J(x,y,t)dx. \tag{9}$$

here, $K(\tilde{y},t)$ is the current density per height transported by the thin conductor at $y = \tilde{y}$ in the bulk $C$. An interpolation function can provide with an estimate for $K(y,t)$ in terms of the $I_k(t)$ constraints for the more general case of each tape carrying a slightly different current than its neighbors. In case of all tapes transporting the same current $I(t)$, $K(y,t) = I(t)/D$ will be constant in the $y$ direction. Hence (9) takes the much simpler form:

$$\frac{I(t)}{D} = \int_C J(x,y,t)dx. \tag{10}$$

The aforementioned constraint can be implemented numerically by introducing only one Lagrange multiplier. The computational time required while using this approach will just now depend upon the mesh density used for discretization of the bulk domain and not on the number of conductors in the original stack. This has the effect of largely reducing the computational time required to simulate large stacks. Moreover, as explained in the following sub-section further speed increase is expected by manually discretizing condition (10).

C. **Homogenization and edge elements**

An even faster computational speed can be obtained by taking advantage of the properties of edge elements. As shown in the appendix, a field $\boldsymbol{H}$ discretized by means of linear edge elements has the following *local* properties:

$$\nabla \cdot \boldsymbol{H} = 0 \tag{11}$$

and

$$\nabla \times \boldsymbol{H} = \boldsymbol{J} \text{ with } \frac{\partial J}{\partial (x,y,z)} = 0. \tag{12}$$

Equation (11) refers to the *local* divergence free of the field $\boldsymbol{H}$, i.e. this condition holds only *within* each element. Edge elements provide continuity for the tangential component in the interfaces between elements. However, discontinuities are allowed for the normal component. In this way, non-divergence free fields can also be represented. For instance see [7,16], both works consider the influence of a magnetic substrate on the AC losses for RABiTS YBCO coated conductors. The magnetic material is modeled by means of a non-linear $\boldsymbol{B} - \boldsymbol{H}$ relation, therefore Gauss law for magnetism $\nabla \cdot \boldsymbol{B} = 0$ no longer enforces $\nabla \cdot \boldsymbol{H} = 0$ as it is the case for linear materials. However, given the aforementioned allowed discontinuity for the normal component, both $\boldsymbol{H}$ and $\boldsymbol{B}$ can be accurately represented in the numerical sense.

The expression in (12) indicates that *within* the element, the current density has a uniform (possibly time dependent) value. Taking advantage of this property, condition (10) can be easily reformulated. For this purpose, let's denote the homogenized bulk equivalent for the stack by $\Omega$, as shown in FIG. 4. Furthermore, let domain $\Omega_i$, a subset of $\Omega$, being discretized by a rectangular structured mesh, so that only one element is used for describing its "thickness". Then, the current density *J within* $\Omega_i$ will be such that $\partial J/\partial y = 0$, while no condition will be imposed on $\partial J/\partial x$. In this way, if the homogeneous bulk domain is divided into $n_s$ subdomains, condition (10) can be represented as a new set of integral constraints:

$$I_i(t) = \int_{\Omega_i} J(x,y,t) dx dy \quad \forall i \in \{1, 2, \ldots, n_s\}. \tag{13}$$

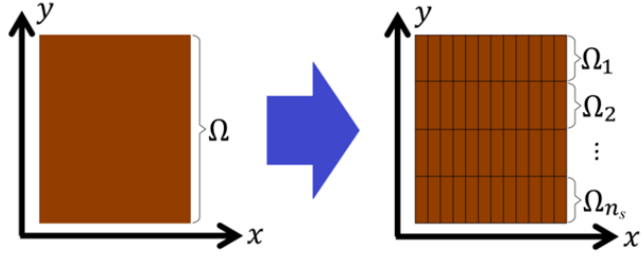

FIG. 4. Discretization of the homogenous bulk domain $\mathbf{\Omega}$ into $\boldsymbol{n_s}$ smaller subdomains.

Although (13) seems similar to (4), note that in general $n_s < n_c$, so that the number of constraints can be arbitrarily reduced at the price of a reduced accuracy with respect to the complete model. As it will be shown in the following sections, excellent accuracy is achieved even with a $n_s$ value several times smaller than the number of tapes in the actual stack modeled. Finally, it is worth noticing that Clem's [8] 3rd assumption that the boundary between the critical and subcritical zones can be approximated with a straight line perpendicular to the tape's surface corresponds to the case of $n_s = 1$.

### III.    TEST CASES FOR VALIDATION

In order to validate the proposed homogenization strategy, the cases of transport current and perpendicular magnetization were considered. A first simulation, following the methodology described in [3], considered a fully featured stack where all layers (up to the μm scale) are included. This meant that all layers shown in FIG. 2 where included in this model. The parameters used were as presented in TABLE I. The second simulation relied on the homogenization technique proposed here. In both cases, stacks of 16, 32 and 64 tapes were considered. For the transport current case, AC currents at 50 Hz were imposed to each of the tapes in the vertical stacks. Amplitudes of the applied currents were 70 A, 60 A and 50 A for the stacks of 16, 32 and 64 tapes, respectively. For the magnetization case, vertical AC magnetic fields at 50 Hz were applied to each stack. Amplitudes of the applied magnetic fields were 90 mT, 100 mT and 110 mT for the stacks of 16, 32 and 64 tapes, respectively.

A Kim like model for the $J_c(\boldsymbol{B})$ dependence of a given IBAD tape at 77K was taken from Thakur et al.[20]. Once multiplied by the volume fraction, the equivalent $J_{c,Eq}(\boldsymbol{B})$ dependence is given by:

$$J_{c,Eq}(\boldsymbol{B}) = \frac{J_{c_0} f_{HTS}}{\left(1 + \frac{\sqrt{k^2 |B_\parallel|^2 + |B_\perp|^2}}{B_0}\right)^\alpha}, \quad (14)$$

where $B_0 = 42.65$ mT, $J_{c_0} = 28$ GA/m², $k = 0.29515$, $\alpha = 0.7$, and $B_\parallel$ and $B_\perp$ are respectively, the parallel and perpendicular components of the magnetic flux density with respect to the tape's surface. These parameters are consistent with 4 mm wide coated conductors with a DC critical current in zero external field of 99.227 A using the $1 \mu V/cm$ critical field criterion.

Structured meshes were used for the original stacks, just as described in [3]. Although other meshes were also considered for the anisotropic bulk, the results presented here correspond to a structured rectangular mesh with 19×50 elements for the whole bulk domain. Triangular meshes were used for the surrounding air domain.

Parameter values used in the simulations are as presented in TABLE I. These correspond to the commercially available 2G HTS tapes manufactured by Superpower [18].

TABLE I. PARAMETER VALUES USED FOR SIMULATIONS.

| Parameter | Value | Description |
|---|---|---|
| $h_I$ | 200 μm | Insulation/Air layer thickness |
| $h_{Cu}$ | 40 μm | Copper layer thickness |
| $h_S$ | 50 μm | Substrate layer thickness |
| $h_{Ag}$ | 2 μm | Silver layer thickness |
| $h_{HTS}$ | 1 μm | YBCO layer thickness |
| D | 293 μm | Unit cell thickness |
| a | 2 mm | Tape half width |
| $\rho_{Ins}$ | 1 Ω·m | Insulation/Air resistivity |
| $\rho_{Ag}$ | 2.70 nΩ·m | Silver resistivity [21] |
| $\rho_{Cu}$ | 1.97 nΩ·m | Copper resistivity [21] |
| $\rho_{Subs}$ | 1.25 μΩ·m | Substrate resistivity [22] |
| n | 38 | Power-law exponent |

## IV. RESULTS AND DISCUSSION
### A. AC Transport Current

The three cases of transport current described in the previous section were simulated. To provide a qualitative comparison between the original stack model and its homogenized counterpart, the case of a stack composed of 32 tapes is analyzed in detail. FIG. 5 shows the magnetic flux density magnitude $|\boldsymbol{B}|$ for half AC cycle in both the original stack of 32 tapes and its anisotropic bulk model representation. The overall profile calculated with the original stack model is well reproduced by the homogenization method at every time step presented. Only a few local differences are seen in the form of horizontal lines in the upper part of FIG. 5. These lines correspond to the actual tapes in the original stack. In the homogenous bulk model, such lines are not observed as the internal layout of the stack has been effectively washed out. The normalized critical current density $J/J_c$ is presented in FIG. 6. To visualize $J/J_c$ in the superconducting layers of the original stack, their thickness was artificially expanded in the vertical direction. Again, the overall profiles in the original stack are well reproduced by the anisotropic bulk model at every time step. Furthermore, it is important to note that since a power law was used for the $E-J$ relation, local overcritical current values are reached in both models. This is expected to have an impact in the critical current of the stack [20].

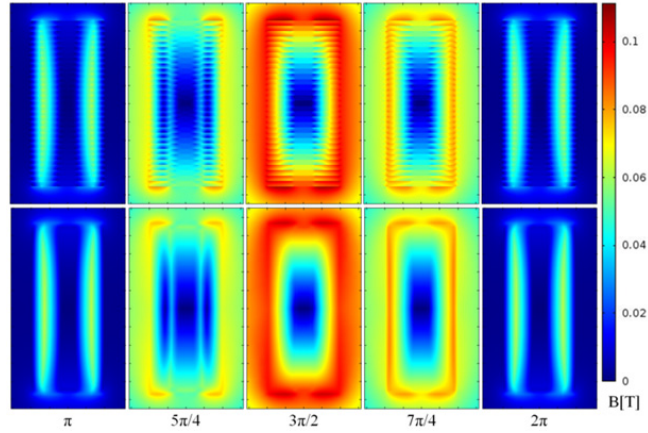

FIG. 5. Magnetic flux density magnitude B [T] for half AC cycle in a stack of 32 tapes in the transport current case. For visualization purposes, domain edges are not plotted. Top: Actual stack. Bottom: Anisotropic bulk model. The actual width of the superconducting layers is 4 mm while the height of the stack is 9.376 mm. The separation between ticks in the plot frames is 1 mm.

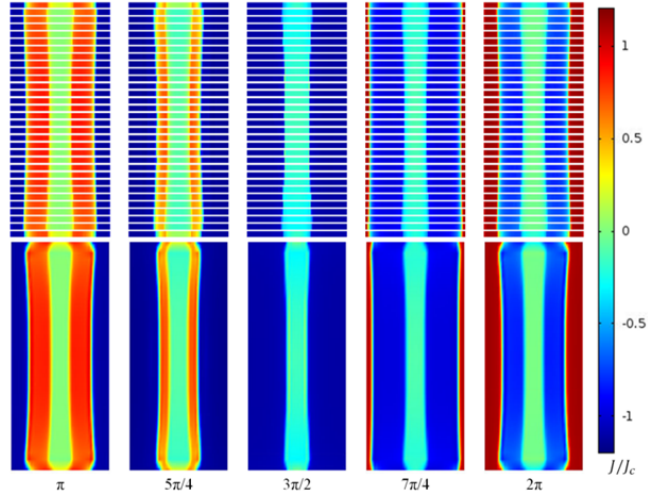

FIG. 6. Normalized current density $J/J_c$ for half an AC cycle in a stack of 32 tapes in the transport current case. Top: Actual stack of tapes. For visualization purposes, the superconducting layers' actual thickness is artificially expanded in the vertical direction. Bottom: Anisotropic bulk model. The actual width of the superconducting layers is 4 mm, while the height of the stack is 9.376 mm.

As seen in FIG. 7, agreement was also found in calculating the instantaneous losses using (6) for a full AC cycle. The first lower "hump" is due to the transient caused by the zero initial conditions for the simulation. Once again, the overall behavior of the original stack is well reproduced by the homogenized model, and only small discrepancies were presented. These could be attributed to the internal geometry simplification and to the fact that the normal conducting layers of the tapes were ignored in the bulk model. Overall, exact reproduction of the results in the original stack model is not likely to happen as information is lost in the homogenization process.

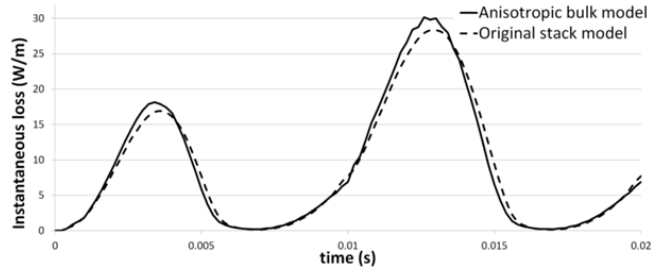

FIG. 7. Instantaneous loss $\xi$ [W/m] for the actual stack of 32 tapes (dashed line) and its anisotropic bulk model (solid line) in the transport current case.

Values for the average AC loss (in W/m) are shown in TABLE II. Calculations made with both the original stack and the homogenized models show good agreement for all the cases considered. Furthermore, the error decreased for the bigger stacks. This could be explained by noticing that with the increasing number of tapes in a stack, the border effects caused by the top and bottom conductors have a lesser impact on the whole stack as the periodic structure dominates. However, since this simple test with three coils does not provide enough data to support a strong conclusion in this matter, further investigations should be latter performed so that other factors such as mesh density are taken into account.

TABLE II. AC LOSSES, TRANSPORT CURRENT CASE

| # of tapes | AC loss (W/m) in original stack | AC loss (W/m) in homogenized stack | Error (%) |
|---|---|---|---|
| 16 | 6.74 | 6.86 | 1.72 |
| 32 | 11.33 | 11.40 | 0.61 |
| 64 | 15.44 | 15.47 | 0.20 |

Computing times for both modeling strategies are shown in TABLE III. Overall, the performance is better for the homogenized model than for the original stack. Furthermore, the speedup increases with the number of conductors considered. Almost two orders of magnitude speedup was achieved for the 64 tapes stack. This is clearly explained by the fact that while more mesh elements – and consequently, degrees of freedom – where needed to simulate the original stack, no mesh increase was needed in the homogenized model. However, simulating even larger stacks with the anisotropic bulk model will likely require more mesh elements to achieve accurate results. Nonetheless, this mesh increase should be expected to have a weak impact in the overall computing time when compared to full scale simulations.

TABLE III. COMPUTING TIME, TRANSPORT CURRENT CASE

| # of tapes | Computing time (s) in original stack | Computing time (s) in homogenized stack | Speedup factor |
|---|---|---|---|
| 16 | 3251 | 639 | 5.09 |
| 32 | 8583 | 472 | 18.18 |
| 64 | 31206 | 426 | 73.25 |

B. Magnetization

FIG. 8 shows the magnetic flux density for half an AC cycle in both the original stack model of 32 tapes and its anisotropic bulk model representation for the applied magnetic field case. Again, the overall profile for the magnitude of the magnetic flux density ***B*** calculated with the original stack model is well reproduced by the homogenization method at every time step presented. Similarly as in the transport current case, one can note that the horizontal lines corresponding to the actual tapes in the original stack model (top part of FIG. 8) are not present in the anisotropic bulk model as the internal layout of the stack has been effectively reduced. Calculated values for the normalized critical current density $J/J_c$ are shown in FIG. 9. Again, as in the transport current case, the profiles in the original stack are well reproduced by the anisotropic bulk model.

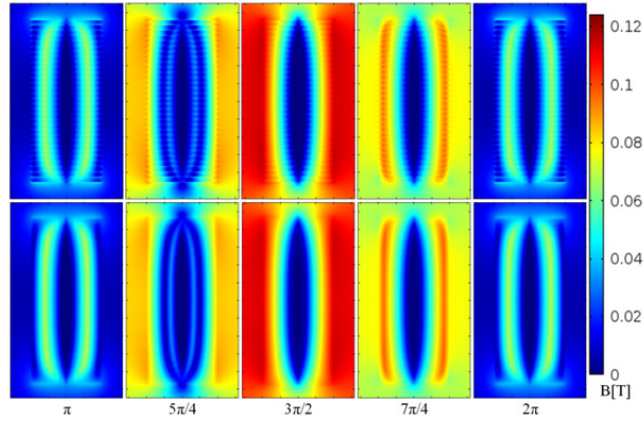

FIG. 8. Magnetic flux density magnitude $B$ [T] for half an AC cycle in a stack of 32 tapes for the magnetization case. A sinusoidal magnetic flux density of 100 mT at 50 Hz was applied vertically to the stack. Results shown at different phase values: from $\pi$ to $2\pi$ in $\pi/4$ increments (from left to right). Top: Actual stack. Bottom: Anisotropic bulk model. The actual width of the superconducting layers is 4 mm while the height of the stack is 9.376 mm. The separation between ticks in the plot frames is 1 mm. For visualization purposes, domain edges are not plotted.

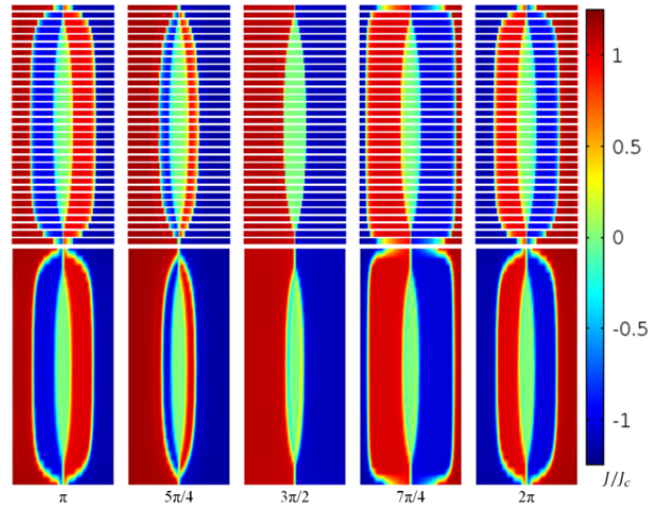

FIG. 9. Normalized current density $J/J_c$ for half an AC cycle in a stack of 32 tapes in the magnetization case. *Top:* Actual stack of tapes. For visualization purposes, only data for the superconducting layers' is plotted. The superconducting layers' true thickness is artificially expanded in the vertical direction. *Bottom:* Anisotropic bulk model. The actual width of the superconducting layers is 4 mm while the height of the stack is 9.376 mm.

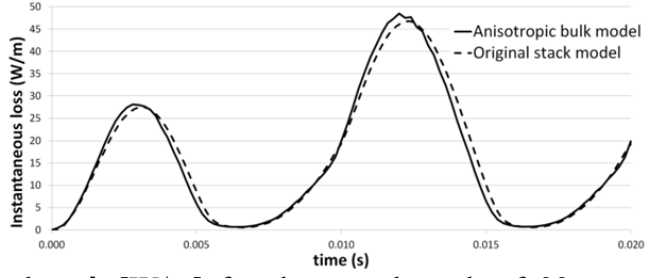

FIG. 10. Instantaneous loss $\xi$ [W/m] for the actual stack of 32 tapes (dashed line) and its anisotropic bulk model (solid line) in the magnetization case.

Good agreement was also found between both calculations for the instantaneous losses (6) for an AC cycle as shown in FIG. 10. Once more, only small discrepancies were seen and results are considered to be good enough to pursue further studies.

Values for the average AC loss (in W/m) for the magnetization study are shown in TABLE IV. Just like for the transport current case, the agreement of the homogenized model with the original stack is good with a decreasing error for bigger stacks.

TABLE IV. AC LOSSES, APPLIED MAGNETIC FIELD CASE

| # of tapes | AC loss (W/m) in original stack | AC loss (W/m) in homogenized stack | Error (%) |
| --- | --- | --- | --- |
| 16 | 10.30 | 10.05 | 2.45 |
| 32 | 19.86 | 19.49 | 1.88 |
| 64 | 41.28 | 40.61 | 1.61 |

Computing times for the magnetization case in both modeling strategies are shown in TABLE V. Again, the performance is better for the homogenized model than for the original stack. Speedup above two orders of magnitude was achieved for the 64 tapes stack.

TABLE V. COMPUTING TIME, APPLIED MAGNETIC FIELD CASE

| # of tapes | Computing time (s) in original stack | Computing time (s) in homogenized stack | Speedup factor |
| --- | --- | --- | --- |
| 16 | 3702 | 539 | 6.87 |
| 32 | 12207 | 647 | 18.87 |
| 64 | 76734 | 676 | 113.51 |

## V. CONCLUSION

In this work we have presented a homogenization method to simulate the electromagnetic behaviour of a stack of HTS coated conductors by means of an anisotropic bulk model. The model uses a continuous $E - J$ relationship, therefore allowing for local overcritical current densities to be considered. Furthermore, the method did not rely on any a priori assumptions for the topology or shape of the critical and subcritical regions.

Both transport current and magnetization cases in stacks of 16, 32 and 64 tapes were considered. The proposed strategy showed good agreement when compared to full 2D simulations performed with the method described in [3]. Calculation of AC losses was performed yielding errors under 2.5% for the 16 tapes stacks. Even smaller errors were obtained for the larger stacks. This is expected as with the increasing number of tapes, the border effects have a lesser impact on the whole stack. In general, the anisotropic bulk model outperformed the full 2D simulations in terms of computational speed. Particularly, a speedup factor of about two orders of magnitude was achieved for the larger coil for both the transport current and magnetization cases.

Considering the speedup achieved, the model presented here can be used for optimization or to obtain important parameters in stack or coil designs such as critical current. For the particular case of generators and motors where a cross section model is used for simulation, the electromagnetic transient behavior of the coils can be calculated without adding a big computational time load [23].

No explicit investigation was carried out to find an optimal mesh distribution, as the speedup was deemed to be significant enough to pursue further modeling and simulation goals.

**ACKNOWLEDGEMENTS**

One of the authors (VMRZ) acknowledges financial support from: Technical University of Denmark, The Research School of the Danish Center for Applied Mathematics and Mechanics, and Vestas Wind Systems.

**APPENDIX: EDGE ELEMENTS**

Edge elements are typically used to represent curl conforming fields while solving PDE's with the finite element method. In what follows, some local properties of the zeroth-order edge elements will be outlined. The treatment presented here follows closely from [24], for further understanding, the interested reader is also pointed to [25]. In the present analysis, both triangular and rectangular elements are considered.

1. **Rectangular edge elements**

Consider the rectangular edge element, as displayed in FIG. 11, used to discretize the magnetic field $\boldsymbol{H} = (H_x, H_y)$. For simplicity and without loss of generality, the element is drawn with its sides being parallel to the coordinate axis. The element has side lengths $l_x$ and $l_y$ and center coordinates $(x_c, y_c)$.

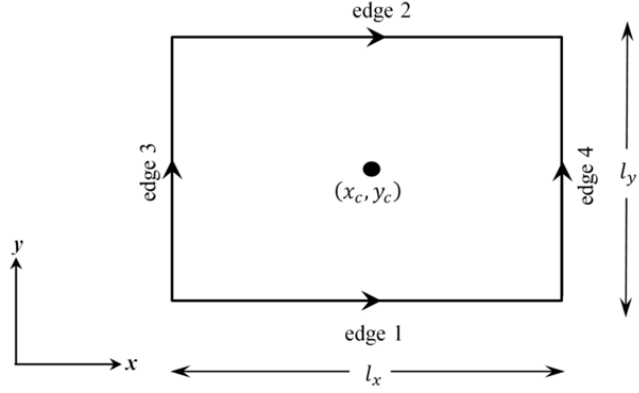

FIG. 11 Rectangular edge element

Assuming that each side of the rectangular element has a constant tangential component assigned, the field within the element can be expressed by the following equations:

$$H_x = \frac{1}{l_y}\left(y_c + \frac{l_y}{2} - y\right)H_{x1} + \frac{1}{l_y}\left(y - y_c + \frac{l_y}{2}\right)H_{x2} \tag{15}$$

and

$$H_y = \frac{1}{l_x}\left(x_c + \frac{l_x}{2} - x\right)H_{y3} + \frac{1}{l_x}\left(x - x_c + \frac{l_x}{2}\right)H_{y4}. \tag{16}$$

Here, $H_{x1}$ and $H_{x2}$ are the tangential components of the magnetic field $\mathbf{H}$ on edges 1 and 2 respectively. In a similar way, $H_{y3}$ and $H_{y4}$ are respectively, the tangential components of $\mathbf{H}$ on edges 3 and 4. Using (15) and (16) one can easily prove that:

$$\nabla \cdot \mathbf{H} = 0 \tag{17}$$

and

$$\nabla \times \mathbf{H} = \left(\frac{(H_{y4} - H_{y3})}{l_x} + \frac{(H_{x1} - H_{x2})}{l_y}\right)\hat{k}. \tag{18}$$

Recalling that the tangential components $H_{x1}, H_{x2}, H_{y3}$ and $H_{y4}$ are space constants and using Ampere's law $\nabla \times \mathbf{H} = \mathbf{J}$, it is easy to see that the current density $\mathbf{J}$ is constant within the element, i.e. $\frac{\partial \mathbf{J}}{\partial (x,y,z)} = 0$.

2. **Triangular edge elements**

A similar treatment can be done in the case of triangular elements. For instance, consider the triangular edge element as displayed in FIG. 12 used to discretize the magnetic field $H = (H_x, H_y)$.

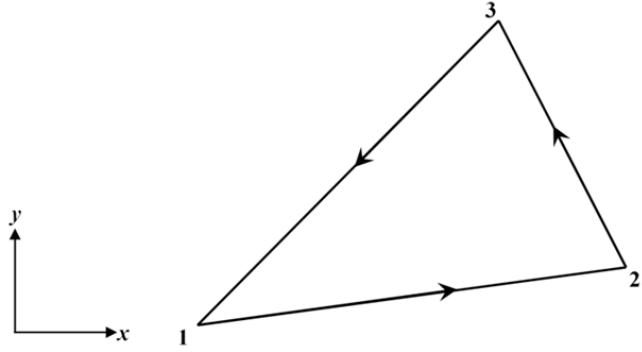

FIG. 12 Triangular edge element

For this purpose, the area coordinates $(L_1, L_2, L_3)$ are considered as follows:

$$L_j(x,y) = \frac{1}{2\Delta}(a_j + b_j x + c_j y), \tag{19}$$

where $\Delta = \frac{1}{2}(x_1(y_2 - y_3) + x_2(y_3 - y_1) + x_3(y_1 - y_2))$ is the surface area of the triangular element and the constant coefficients $a_j$, $b_j$ and $c_j$ are given by:

$$\begin{aligned} a_1 &= x_2 y_3 + y_2 x_3 & b_1 &= y_2 - y_3 & c_1 &= x_3 - x_2 \\ a_2 &= x_3 y_1 + y_3 x_1 & b_2 &= y_3 - y_1 & c_2 &= x_1 - x_3 \\ a_3 &= x_1 y_2 + y_1 x_2 & b_3 &= y_1 - y_2 & c_3 &= x_2 - x_1 \end{aligned} \tag{20}$$

Here, the $x_j$ and $y_j$ values refer to the coordinates of the vertices of the triangle in FIG. 12. Now, let's consider the following vector functions:

$$\mathbf{N}_1 = (L_1 \nabla L_2 - L_2 \nabla L_1) l_{12}, \tag{21}$$

$$\mathbf{N}_2 = (L_2 \nabla L_3 - L_3 \nabla L_2) l_{23}, \tag{22}$$

and

$$\mathbf{N}_3 = (L_3 \nabla L_1 - L_1 \nabla L_3) l_{31}, \tag{23}$$

where $l_{ij}$ is the length of the edge joining the vertices $i$ and $j$ as presented in FIG. 12. Then, the magnetic field $\mathbf{H}$ *within* the element is given by:

$$\mathbf{H} = \sum_{i=1}^{3} \mathbf{N}_i H_i, \tag{24}$$

where $H_i$ is the tangential component of the magnetic field on the $i$th edge. Just like in the previous section, it is easy to see that:

$$\nabla \cdot \boldsymbol{H} = 0 \tag{25}$$

and

$$\nabla \times \boldsymbol{H} = \frac{1}{\Delta}\sum_{i=1}^{3} H_i l_i \hat{k}. \tag{26}$$

Again, recalling that the tangential components $H_i, i \in \{1,2,3\}$, are space constants and using Ampere's law $\nabla \times \boldsymbol{H} = \boldsymbol{J}$, it is easy to see that the current density $\boldsymbol{J}$ is constant within the element, i.e. $\frac{\partial \boldsymbol{J}}{\partial (x,y,z)} = 0$.

It is easily seen that for both triangular and rectangular meshes, use of zeroth-order edge elements to discretize the magnetic field yields the following *local* properties:

$$\nabla \cdot \boldsymbol{H} = 0 \tag{27}$$

and

$$\nabla \times \boldsymbol{H} = \boldsymbol{J} \text{ with } \frac{\partial \boldsymbol{J}}{\partial (x,y,z)} = 0. \tag{28}$$

Furthermore, from equations (16) and (28), it is easy to see that use of zeroth-order edge elements allows for a direct computation of the current density from the calculated magnetic field components without the need for additional numerical differentiation, hence providing with a high degree of accuracy.